\documentclass{svmult}
\usepackage{type1cm,graphicx,newtxtext,newtxmath,csquotes,xcolor}
\usepackage[bottom]{footmisc}
\usepackage[backend=biber,style=biblatex-spbasic,sorting=nyt,natbib=true,uniquename=init,dashed=false]{biblatex}
\bibliography{ArenhartArroyoMelo.bib}
\begin{document}

\title*{Disentangling scientific realism from anti-exceptionalism\thanks{Invited contribution to appear in \fullcite{molick202x}. Submitted: November 27, 2023; Accepted: August 4, 2024.}}
\author{}
\author{Jonas R. B. Arenhart \and Raoni Arroyo \and Ederson Safra Melo}
\institute{Jonas R. B. Arenhart \at Department of Philosophy. Federal University of Santa Catarina. Florianópolis, Brazil. \at Graduate Program in Philosophy. Federal University of Maranhão. São Luís, Brasil.
\at Research Group on Logic and Foundations of Science (CNPq).
\at Research Group on Philosophy of Logic and Language (GFILL/CNPq).
\at email{jonas.becker2@gmail.com}
\and Raoni Arroyo \at Centre for Logic, Epistemology and the History of Science. University of Campinas. Campinas, Brazil. \at Support: grant \#2021/11381-1, São Paulo Research Foundation (FAPESP).
\at Research Group on Logic and Foundations of Science (CNPq).
\at Research Group on Philosophy of Logic and Language (GFILL/CNPq).
\at \email{raoniarroyo@gmail.com}
\and Ederson Safra Melo (Corresponding author) \at Department of Philosophy, Federal University of Maranhão, São Luís, Brasil.
\at Research Group on Philosophy of Logic and Language (GFILL/CNPq).
\at \email{saframelo@gmail.com}}

\motto{ }

\maketitle

\abstract{~Scientific realism is, currently, one of the most well-entrenched background assumptions of some relevant versions of anti-exceptionalism about logic. We argue that this is a sort of sociological contingency rather than a metaphilosophical necessity. Drawing parallels with the metaphysics of science (as applied to quantum foundations), we try to bring the realist assumptions of anti-exceptionalism to light, to demotivate the necessary connection between realism and anti-exceptionalism, briefly exploring the possibility of adopting antirealism as the background default view of science instead.}

\abstract*{~Scientific realism is, currently, one of the most well-entrenched background assumptions of some relevant versions of anti-exceptionalism about logic. We argue that this is a sort of sociological contingency rather than a metaphilosophical necessity. Drawing parallels with the metaphysics of science (as applied to quantum foundations), we try to bring the realist assumptions of anti-exceptionalism to light, to demotivate the necessary connection between realism and anti-exceptionalism, briefly exploring the possibility of adopting antirealism as the background default view of science instead.}

\keywords{Anti-exceptionalism about logic,
Logical epistemology,
Metaphilosophy,
Philosophy of logic,
Realism and antirealism}

\section{Introduction}\label{sec:1}

It is an undisputed fact that there is an infinity of distinct formal systems of logic. They are all presented through the use of various mathematical formulations, either in terms of a semantic approach, or else in proof-theoretic versions, and, in a sense, these systems differ in the answer they provide for which inferences are considered to be licit. Facing a problem where logical tools are required, we find ourselves having to choose an appropriate logical theory. One such context, according to philosophers of logic, is quite special, it involves addressing the substantial problem of what \textit{actually} follows from what, the so-called \textit{canonical application of logic}; in this case, we are in the business of having to choose the \textit{correct} account of logical consequence for inferences drawn in natural language. That is, we have to spot which among the many systems available correctly describes what follows from what in an extra-systematic sense \citep[see][chap.~2]{Haack1978-HAAPOL-2}. So, this view assumes that the correct account of logical consequence corresponds to some informal, or extra-systematic, conception of logical consequence. As \citet{martin-2021} puts it:
\begin{quote}
Probably the most prominent and philosophically interesting example of such theory choice is over logics that aim to not just solve a certain technical or philosophical problem, but aim to provide a general account of validity---\textit{what propositions follow from what}. Multiple logics have been proposed as viable candidates to capture validity: paraconsistent logics, intuitionistic logics, paracomplete logics, quantum logics, and of course classical logic. Given that these various logics license different rules of implication, often with significant repercussions, it clearly matters which logic we ultimately endorse. \citep[9070]{martin-2021}.
\end{quote}
One frequently finds the claim in current philosophy of logic literature that, facing this specific version of the problem of theory choice, the traditional accounts of the epistemology of logic just do not seem to be up to the task. Traditional epistemology of logic is said to rely on \textit{a priori} methods---rational intuition or semantics of logical vocabulary---that, besides being problematic in themselves, just cannot account for the fact that logic is now thought to be revisable, i.e. that we may change our view as to which logical theory correctly describes validity depending on the evidence we have at our disposal at a given time \citep[see][]{priest-2016,hjortland-2017,martin-2021}. As a result, logic, taken as a discipline, lacks an appropriate epistemology.

But not all is lost. There is a new tendency in philosophy of logic, going under the name of \textit{anti-exceptionalism about logic} (AEL, hereafter), that tries to fill this gap by bringing logical knowledge closer to the knowledge of empirical sciences. AEL is the view that logic is not `exceptional' (as the name claims) with respect to other sciences; that means, among other things, that there is a kind of `common epistemology' for logic and for empirical sciences. According to the oft quoted characterization of AEL:
\begin{quote}
Logic isn't special. Its theories are continuous with science; its method continuous with scientific method. Logic isn't a priori, nor are its truths analytic truths. Logical theories are revisable, and if they are revised, they are revised on the same grounds as scientific theories. These are the tenets of anti-exceptionalism about logical theories \textelp{}. \citep[632]{hjortland-2017}.
\end{quote}
It is not an accident that in this epistemological proposal, the focus has been put on logical theory choice. Particularly, the idea is to bring the methodology of scientific theory choice to logic as an attempt to boost the latter's epistemology. The plan is quite simple: science has an incredible track record of success; let logic benefit from that as well! By using a method of theory choice similar to the one employed in science we not only recognize that logical theories are revisable,\footnote{~Which, according to some, would also account for the failure of the \textit{a priori} character and the analyticity of logic; however, see \citet{BeckerArenhartERK,ArenhartCunha2023} for further discussion.} but also employ some theory choice methods whose reliability in delivering the goods has an incredible success rate \citep[see][]{priest-2014,priest-2016,hjortland-2017,martin-2021}.

In this paper, we explore one specific aspect of AEL's proposal of addressing the idea of continuity between logic and science in terms of the continuity in methods of theory choice or theory appraisal in general. Our major claim is that AEL---as characterized in most of current literature---takes for granted that there is a unified and completely consensual approach to what are the methods and aims of science. This supposition results in AEL subscribing to a straightforward realist description of the workings of science.\footnote{~The same claim is also advanced by \citet{Erickson2021}, although we shall pursue it here in different lines.} Given this view, it would be just a matter of transferring the use of those methods to logic, and at least part of the seeming stalemate of theory choice in the philosophy of logic would be solved. But one can never forget that realism is not the only way to frame the aims and proper understanding of science.

As we shall suggest, continuity with science can be approached through different and incompatible lines, with the proper aims and methods of science being a highly disputed topic in the philosophy of science literature. As a result, the idea of continuity between logic and empirical science is not a simple matter, and framing it as a question of theory choice by use of abduction or some focus on the explanatory power of theories is just one possibility among many. It is not to be expected that all of those different ways to frame the similarity between science and logic would be congenial to the aims of actual major formulations of AEL. That, by itself, is not an argument against AEL, but against taking for granted that there is only one way to link logic with other branches of science. As even some anti-exceptionalists are now recognizing, the characterization of AEL in terms of abduction and logical theory choice faces difficult challenges; our hope is that by bringing some of the underlying commitments of the view to light, more appropriate approaches may be sought. 

The structure of this chapter is as follows. In section \ref{sec:2}, we address the question of which view of science AEL seems to be taking for granted. In section \ref{sec:3}, we borrow some instructive illustrations from the case of quantum mechanics and the metaphysics of science, where metaphysics tries to obtain some epistemic warrant from the success of science. The case is instructive because it reflects a common pattern in attempts of naturalization or approximations with science in general. In section \ref{sec:4}, we bring to light some general lessons that AEL may learn from the previous discussions. We conclude in section \ref{sec:5}. 

\section{AEL and logical realism}\label{sec:2}

Let us begin by putting it again in explicit terms: AEL, as a general thesis, advances the claim that logic should have some kind of continuation with science, being able to join science in its successful achievements by using the same kind of methodology. As remarked, that claim concerning continuity is very flexible, given that the nature of such continuity may be explained in many distinct ways. All would be very unproblematic if there were a unified and consensual methodology of science, recognized by everyone and applied across the board. But things are not like that \citep[see in particular the discussion about methodology in][]{Bueno2017}. 

In order to put those issues in clearer terms, we begin, in this section, by presenting AEL with some of its nuances. Even though there is no uniformity in the distinct characterizations of AEL, some of the formulations found in the literature have basic common threads that will allow us to develop our point \citep[for a criticism of the ambiguity in the characterization of AEL, see][]{Rossberg2021}. Again, recall that one of the main targets of AEL is the alleged weaknesses of the traditional epistemology of logic, which should be overcome by advancing a science-based epistemology, focused on theory choice. 
\begin{quote}
Motivated by weaknesses with traditional accounts of logical epistemology, considerable attention has been paid recently to the view, known as AEL, that the subject matter and epistemology of logic may not be so different from that of the recognised sciences. One of the most prevalent claims made by advocates of AEL is that theory choice within logic is significantly similar to that within the sciences \citep[285]{MartinHjortland2021}.
\end{quote}
The backdrop is to substitute current inappropriate epistemology (mainly thought to be \textit{a priori}, recall) with the quite tested and approved epistemology of science. The focus lies, as the quote makes clear, primarily in theory choice, related to the claim that there are no---or there \textit{should} be no---significant differences between science and logic on these matters. In this scenario, \citet[286]{MartinHjortland2021} set up a \textit{Methodological version of AEL} in the following terms: ``Theory choice within logic is similar in important respects to that of the recognised sciences''. So, according to this perspective, logical laws and inferences are justified---and perhaps come to be known---on the basis of theory choice, whose methodology is similar to the one employed by ``recognised sciences''. 

The problem then is: how does theory choice work in the recognised sciences? In the methodological AEL camp, \textit{abductivism} is often taken as the proper method for theory choice in the sciences, and logic should follow the same procedure \citep[see][]{priest-2014,priest-2016,hjortland-2017,martin-2021}. According to this view, logical propositions aren't justified directly on the basis of intuitions or semantic definitions (as the traditional accounts of logical epistemology would have it), but rather by being part of a logical theory that can be chosen as the best explanation of the available logical evidence. Roughly, one theory will be selected as the best explanation, mainly, by its ability to best accommodate the relevant data (which, remember, is supposed to be \textit{a posteriori}), and also by taking into account a number of theoretical virtues, such as explanatory power, simplicity, avoidance of \textit{ad hoc} elements, consistency \citep[see][32]{priest-2016}.

That is not all to the methodological approximations, however. In the face of some problems with abductivism in logic,\footnote{~See \citet{Hlobil2020,BeckerArenhartERK}.} \citet{MartinHjortland2021,MartinHjortland2022} have proposed a distinct formulation of AEL, arguing that logical theories are not chosen for their ability to fit the data, but rather by their ability to make successful predictions. According to \citet[p.148]{MartinHjortland2022}, this view, called \textit{logical predictivism}, is the ``most detailed and plausible version of methodological AEL available''. In this view of logic, there are specifically logical phenomena that the logical theories try to explain, and the fruitfulness of these explanations is judged by the successful predictions \citep[288]{MartinHjortland2021}. 

Despite their seeming differences, logical abductivism and logical predictivism have some common background assumptions. In both instances, the anti-exceptionalist aims to justify the correctness of a particular logical system by emphasizing a key feature of that system, one that sets it apart from rival systems that either lack it or possess it to a lesser extent. The abductivist wishes to conclude that a system is the correct one because it suits the evidence, while the predictivist wishes to argue for a system because it is taken to correctly predicts what is actually valid (remember also Ben Martin's quote at the beginning of this chapter). 

What lies behind both approaches, we suggest, is basically a broadly realist conception of logic and of science.\footnote{~For further critical discussion relating claims that abduction and scientific realism are part of the same piece, see \citet[chap.~2.3]{vanfraassen1980} and the references therein} The idea that there \textit{is} a matter of fact about logical consequence, and that a common uniting feature behind the abductivist and the predictivist approaches is to argue that different theories try to describe or explain such facts. This is found in most discussions of what the data of logic could be (semantic data, or extra-linguistic data), and the view that the business of logic is description (and not, or not mainly, prescription). Among others, Graham Priest puts the point as follows, distinguishing a logical theory (\textit{logica docens}) from the logical facts (\textit{logica ens}): 
\begin{quote}
\textelp{} it is crucial to distinguish between logic as a theory (logic docens, with its canonical application), and what it is a theory of (logica ens). In the same way we must clearly distinguish between dynamics as a theory (e.g., Newtonian dynamics) and dynamics as what this is a theory of (e.g., the dynamics of the Earth). \citep[215]{priest-2014}.
\end{quote}
And he further explains what \emph{logica ens} consists of as follows:
\begin{quote}
These are the facts of what follows from what---or better, to avoid any problems with talk of facts: the truths of the form `that so and so follows from that such and such'. \citep[220]{priest-2014}.
\end{quote}
According to the views proposed, then, logic, given that it simulates empirical science, is attempting to find out which among the many options available is the correct description of the notion of validity, or validity \textit{simpliciter}. Both logic and science, on this view, are looking for the truth about a certain independently existing phenomenon, where truth is understood in a kind of correspondential sense (\textit{viz.} discussions concerning `the one true logic'). That is, this is an openly realist account of the goals and methodology of logic. Realism is a very difficult position to characterize, but, still, we can find the following commonality among the diversity: 
\begin{quote}
Amidst these differences, however, a general recipe for realism is widely shared: our best scientific theories give true or approximately true descriptions of observable and unobservable aspects of a mind-independent world. \citep{chakravartty-sep-scientific-realism}.
\end{quote}

Now, there is no question that it is perfectly fine to try to be a realist in logic. Depending on how the term is characterized, logical monists are realists of a sort, and perhaps most logical pluralists are realists too (\textit{i.e.} when the characterization of these terms involves the idea of one or more than one correct or true description). But what cannot be forgotten is that realism is not the only game in town when it comes to the philosophical understanding of empirical science. As a result, it cannot be assumed without further ado that the \textit{only} way to bring logic closer to science is by means of adhering to a realist description of the efforts of science. Of course, one may argue that realism is the \textit{best} way of doing so, but then this should come up as a conclusion, not a premise. As \citet[1850]{guaypradeu2020} diagnosed, a similar attitude was taken in past discussions on metametaphysics concerning the relationship between science and metaphysics; as they called it, this was an ``oversimplification''  that ``tended to reduce science-based metaphysical approaches to realism''. Something similar, we claim, is happening to the efforts offered by AEL to bring logic closer to science. 

We seek a parallel in the metaphysics of science, for we think there's a lesson to be learned from it. By avoiding a sort of restricted account of science, one opens up the scope of approaches to the understanding of science and, as a result, further possibilities that could help us to enlighten the understanding of logic are also now available. Why not, for instance, think that logic could be approached through the lights of distinct, not straightforward realistic, approaches to the philosophy of science? If one is on the search for approximating science and logic, one is also required to look for antirealist accounts of science. Of course, scientific antirealism comes in many distinct versions, and making one of those explicit would certainly open the way for distinct forms of AEL and antirealist approaches to logic. We are not claiming that they would all be equally interesting, but the fact that there are other options on the table makes the choice for a realist approach hardly obvious, to say the least.

One way of seeing how this could be done would be to adopt a form of constructive empiricist approach to logic---or, at least, to consider some of the features that constructive empiricism uses to approach science, and approach logic with the same tools. According to constructive empiricists, we should not believe in the truth of our scientific theories when it comes to the unobservable aspects of the description of reality; rather, given the evidence available, only belief in the empirical adequacy is warranted (van Fraassen \citeyear{vanfraassen1980,vanfraassen1989}). That would lead us to a different picture of the task of an epistemology for logic that connects with science. In fact, that is a dimension worth exploring a little bit, if by way of illustration only, of how an understanding of logic could benefit from distinct understandings of science. 

According to the empiricist version of science, the existence of multiple theories, all equally compatible with the evidence, is a good reason for us to be agnostic about the literal truth of such theories (that is the moral of the famous argument from the underdetermination of the theories by the data, to be further explored in section \ref{sec:3}). A similar story could be provided in the case of logic by the fact that there are multiple logical frameworks, provided that they can give us the same important deductive results in many of the most important fronts (as some of them do with recapture strategies; see also our discussion in \ref{sec:4}). So, just as distinct empirical theories compatible with the evidence would be telling us distinct ways the world could be, distinct systems of logic would be delivering for us distinct ways the logic behind our inferences could be (many-valued, constructive, inconsistency-tolerating, and so on), if each of such logical frameworks were true \citep{vanfraassen1989}.\footnote{~For alternative formulations of AEL not focusing on logical theory choice and not openly realistic, see for instance \citet{Arenhart2022Theoria,Tajer2022,CommandeurForthcoming}.} What we would get from such a multiplicity of theories would be a better understanding of logical consequence, even if none of the candidates could be actually known to be `the' true one. 

For an illustration of how such an approach would benefit from actual evidence from science, we shall consider current debates concerning different levels of theory choice in quantum mechanics. 

\section{The quantum case}\label{sec:3}

One of the crucial claims of AEL is related to the respectability, on epistemic grounds, that logic would derive from employing the same methodology of theory choice as the one employed by empirical sciences. As we have suggested in the previous section, the way the story is told in current literature, one is quickly led to a realist account of both science and logic, with the choice of a theory by abduction leading one to consider the chosen theory as a candidate to be the true theory. 

Such a claim, however, rests on many unsettled issues. The most pressing one is that realism about science faces many difficult challenges, being diverse versions of underdetermination among the most pressing issues. Let's get specific about this by analyzing one of the favorite examples for philosophers of science: (non-relativistic) quantum mechanics. Quantum mechanics is the perfect medium for the realist/antirealist debate because it exemplifies very well what's at stake on both epistemic stances. On the realist side, it showcases undeniable empirical success. Quantum mechanics is outstandingly well-confirmed, so it enjoys all the empirical success a scientific theory can dream of. On the antirealist side, it showcases underdetermination, because quantum mechanics exemplifies such cases in so many distinct levels. Here we bring four such levels to the fore, respectively: dynamical, ontological, metaphysical, and logical levels. Let's briefly see how this plays out, and how an ensuing antirealist approach to logic may be motivated in this context too.

In familiar quantum-mechanical situations, one may describe the state of a quantum system (\textit{e.g.} an electron) as a sum of its possible states. That is to say that, (suppose) if an electron $e$ could assume the state of being located at $A$ and thus assume the state-vector $|A\rangle_e$; and if it could \textit{also} assume the state of being located at $B$ and thus assume the state $|B\rangle_e$, the correct way of describing the electron in such an event is the sum $|A\rangle_e+|B\rangle_e$. This vector sum is called ``superposition'', and the field of quantum foundations spends lots of effort in spelling out whatever physical meaning it could amount to as it cannot be, for instance, meant that the electron is located in region $A$, nor in region $B$, nor in both $A$ and $B$ at the same time, and not in any of these places---which is the four logical possibilities we have of describing ordinary objects in our experience \citep{albert1992}. However, it turns out that $|A\rangle_e+|B\rangle_e$ is the empirically adequate way of describing electrons in a wide array of experiments ranging from quantum optics to Nobel-prize winning quantum correlation experiments \citep{lewis2016}. Something must be going on there, and quantum foundations is the field that (among other things) is supposed to explain precisely what is it. The usual way of presenting what's wrong with the above picture goes something like this, in an either-or situation that started out as a dilemma \citep{bell1989}, evolved into a trilemma \citep{maudlin1995}, and now it's better described as polylemmas \citep{muller2023}:

\begin{itemize}
    \item The superposition is couched in the wrong terms. If the dynamics led us into a superposition situation, it should be plainly wrong as we never get to observe superposition---not in the physics lab, not in daily life. There must be \textit{another} kind of dynamics that gives us a determinate position (to stick with the example), and this dynamics is often called the ``collapse''. So there's a dynamics for superposition, and another one to get rid of it. This is often called standard quantum mechanics, or simply ``quantum mechanics''.
    \item The superposition is couched in incomplete terms, qua epistemic, so that the position of the electron is always determinate albeit we don't have epistemic access to it. So the dynamics is often said to be one of hidden variables, or hidden parameters, that specifies the determinate electron's position at all times. The way to go for it is \textit{e.g.} Bohmian mechanics.
    \item The superposition is read literally: both states describe what is actually going on, albeit not in the same world. No need to add hidden variables or other dynamics: the dynamical law that describes the physical situation as a superposition is everything and it is right. This is the Everettian quantum mechanics way to go.
\end{itemize}

This is how far science goes, delivering us the so-called `quantum interpretations'. What is illustrated is the first level of underdetermination (of theory by evidence), and there's no \textit{scientific} reason to choose one among the options offered. This is to say that physics alone does not determine which is the right way to go. Notice that this does not mean that absolutely everything goes, as the interpretations must be empirically adequate to current evidence. To stress our methodological point, however, the following must be clear: QM does not determine its interpretation.

In the remainder, we'll stick with the first case in the textbook story mentioned before, the one of standard quantum mechanics. Recall that no \textit{scientific} reason compels one to do so; we're picking this one for the sake of building our argument.

So let's keep digging. The ontological level deals with what exists according to each interpretation. Recall that we are dealing with a single interpretation, and there's no consensus here as well. For instance, one might put forth the idea according to which the furniture of the world is composed of (among other things) particles in three-dimensional space and human consciousness that cause the collapse dynamics \citep{debarros-oas-2017}; alternatively, one can \textit{also} point out that there are no particles or minds, and what exists in the fundamental level is a high-dimensional field called the wave function \citep{ney2021}. Surely there are more alternatives than that, but two will suffice. This is the second level of underdetermination, and---just as it occurred with the first level---there's no \textit{scientific} reason to choose among them. It doesn't matter which one we pick for us to keep digging beyond this point, as long as we keep committed to our first choice, \textit{viz.} standard quantum mechanics.

Enter the metaphysical level. Due to the indistinguishability postulate, the permutation of quantum objects does not give rise to a new physical situation \citep{huggett-imbo2009}. Given any two physical systems, say, $|\psi\rangle$ and $|\phi\rangle$, their permutation of their tensor product $\otimes$ results in the equality $|\psi\rangle\otimes|\phi\rangle=|\phi\rangle\otimes|\psi\rangle$. This is often contrasted with the so-called ``classical'' case, which can be illustrated with a statistical example. Suppose one wants to arrange two particles, $1$ and $2$ inside two boxes, $A$ and $B$; the four classical possibilities ($C_1$--$C_4$) to do so are:
\begin{description}
    \item[$C_1$. $A^1A^2$:] Both particles are inside box $A$;
    \item[$C_2$. $B^1B^2$:] Both particles are in box $B$
    \item[$C_3$. $A^1B^2$:] Particle $1$ is inside box $A$, and $2$ in $B$;
    \item[$C_4$. $B^1A^2$:] Particle $1$ is inside box $B$, and $2$ in $A$.   
\end{description}
The statistical weight for each one of these possibilities is $\tfrac{1}{4}$, and this is called the Maxwell--Boltzmann statistics.

So far, so good, but QM disrupts that with the indistinguishability postulate. Notice that the third and fourth cases differ only in particle permutation, so the statistics cannot generate a new physical situation. Here's what happens, then, with the quantum possibilities ($Q_1$--$Q_3$):
\begin{description}
    \item[$Q_1$. $A^{\bullet\bullet}B^{\circ\circ}$:] Both indistinguishable particles are inside box $A$;
    \item[$Q_2$. $A^{\circ\circ}B^{\bullet\bullet}$:] Both indistinguishable particles are inside box $B$;
    \item[$Q_3$. $A^{\bullet\circ}B^{\bullet\circ}$:] One indistinguishable particles is inside each box, $A$ and $B$.
\end{description}
$Q_1$--$Q_3$ have the same statistical weight. Notice how $Q_3$ replaces cases $C_3$ and $C_4$. This is called the Bose--Einstein statistics.

The above-mentioned change in the statistics for the classical and quantum cases is often explained by a change in the \textit{metaphysics} of quantum objects. In the traditional way of thinking about this situation, it is thought that, whereas classical objects are endowed with \textit{individuality}, quantum objects are not. This would explain why $C_3$ and $C_4$ are different physical situations, and why they all collapsed into $Q_3$: quantum objects have no individuality, so it doesn't matter if one exchanges them. Quantum objects, in this view, are non-individuals \citep{krause-arenhart-bueno2022}. Whereas classical objects can be cashed out in terms of haecceities (the non-qualitative property of ``being identical with itself'', quantum objects should be understood by their lack of haecceity. This is the Received View of quantum (non-)individuality \citep{frenchkrause2006,arenhart2017}. Quantum objects, however, can \textit{also} be cashed out in haecceistic terms---just as their classical cousins---if one wishes to do so \citep{french2019}. One might simply point out that haecceities are spelt out in substantialist terms, or simply posited as primitive \citep{morganti2015}.

The methodological point that we wish to stress with all this is the following: \textit{none of the above-mentioned metaphysical views for quantum objects can be settled by QM itself.} This is underdetermination level three: the ``underdetermination of metaphysics by physics'' or simply ``metaphysical underdetermination'' \citep{frenchkrause2006}. The individuality/non-individuality profiles may always be added as a metaphysical maneuver to further interpret quantum objects \textit{qua} objects without changing physical/empirical aspects of quantum mechanics. Because of that, the physical/empirical domain cannot tell which of such metaphysical `packages' is more 
appropriate/true. At the end of the day, metaphysics of science is still metaphysics---and the fact that it is applied to science doesn't make much progress in terms of theory choice \citep{arroyo-arenhart-2022synt}.

On to logical underdetermination at level four. Take the case for quantum statistics, specifically $Q_3$. The particles inside each box, $A$ and $B$, are \textit{absolutely indistinguishable}. One way of interpreting it is to state that quantum objects lack the notion of ``sameness'' \citep[to use the words of][197]{schrodinger1998}, and another way is stating that they don't.

In logical terms, it has been suggested that it all comes down to whether or not the reflexivity of identity---understood in the terms of $\forall x(x=x)$---is being violated by quantum mechanics. If not, classical logic is everything one needs in this matter, but then one must do some tricks to account for absolute indistinguishability (see \cite{frenchkrause2006,krause-arenhart2018} and \cite{bigaj2022} for discussion). If reflexivity of identity is being violated by $Q_3$, then one must jump right into some non-classical system. There's one tailored made for this case: Schrödinger logics \citep{dacosta-krause1994} and its set-theoretical level through the use of quasi-sets \citep{krause1992,french2019}. Roughly speaking, in such non-reflexive logics the expression $x=x$ is a well-formed formula only in specific cases, \textit{viz.}, cases in which the objects in question are not quantum objects, but so-called ``everyday'' ones (such as tables, chairs, and so on). For quantum objects, identity (``$=$'') doesn't apply; for everyday objects it does. For the former, quasi-set theory offers a weaker relation, the one of indiscernibility (``$\equiv$'').

What is important for us now, however, are not the details of how such theories would work, but rather, that non-reflexive logics, in the form of quasi-set theory, seem to encompass a metaphysics of non-individuality. In this sense, they rival classical logic approaches to identity, which, it is claimed, codify individuality right from the start \citep[see][chap.~6]{frenchkrause2006}. The trouble is that with the underdetermination of the metaphysics by the physics, we are stuck also with no additional evidence for the choice of the appropriate logic, classic or non-reflexive. In fact, both classical set theory and quasi-set theory can provide for the mathematical frameworks needed to develop quantum mechanics, so that there is no mathematical or physical fact of the matter that could help us to choose one of the logics.\footnote{~See \citet{Domenech2008} for a development of Fock spaces in quasi-set theory.} The result is deep underdetermination also for logic.\footnote{~See \citet{ArenhartArroyo2023Road} for discussion on these levels of underdetermination.} 

Now, notice that all these kinds of underdetermination happen in the \textit{same domain} of investigation. It is not as if we're having trouble unifying the dynamics/ontology/metaphysics/logics of \textit{e.g.} non-relativistic quantum mechanics and special relativity, nor quantum mechanics and pediatrics. A pluralist account may handle that case more quickly, stating that each domain has it own specifics \citep[see, for example,][]{bueno2010,ruetsche2015,Bueno2017}. \citet{bueno2010}, for instance, argues that we should not bother deciding between the logics of quantum mechanics \citep[the so-called ``quantum logic''\footnote{~We should mention in passing that quantum logic, like quasi-set theory, was tailored-made for the quantum case; unlike quasi-set theory, however, quantum logic doesn't address specific questions of identity and indiscernibility.} presented by][]{birkhoff-vonneumann1936} and logics of other domains; \citet{ruetsche2015} has a similar move by stating that we should peacefully accept the plurality of theories (specifically dynamics and ontology) for different domains. The cases we've brought to the table are even more radical, \textit{viz.}, of quantum mechanics itself having no means whatsoever to decide these matters on purely quantum-theoretical grounds. 

As we have seen, the appeal to continuity with science (whatever that means exactly) was not enough to settle the four above-mentioned levels of underdetermination. Physics is deadly silent on such matters, being compatible with every single one of the aforementioned possibilities. This means, among many other things, that quantum mechanics has no clear method for theory choice. So when it comes to the matter of choosing a correct logical theory concerning some aspect of ``logical facts'', quantum mechanics is not your ally. The moral of this example is what went wrong: AEL shouldn't be so quick in hoping that science is always teaching us clear lessons about the true theory to be chosen. Underdetermination is a widespread phenomenon, and the antirealist makes a lot of it.\footnote{~This is not to say that underdetermination is a knockdown argument against scientific realism, though. Alternatives are available. To mention a few, it has been proposed that: we should leave quantum mechanics out of the scientific realist picture \citep{hoefer2020}; we should explicitly use extra-empirical virtues to theory choice in these cases \citep{callender2020}; that there are commonalities between underdetermined options that must be sought \citep{french2014,fraser-vickers2022}. All of them compromise the scientific realist starting goal a little bit, but that's not why we brought up underdetermination arguments. Realist's attitude towards underdetermination be as it may, the point we wish to stress is that theory choice in science is far from being unproblematic.}

\section{The lessons for AEL}\label{sec:4}

As we have illustrated with the case of quantum mechanics, there are at least some reasons to believe that an antirealist account of science should also be considered when issues related to the continuity between logic and science are taken into account. As a result, those looking for inspiration in science could also attempt to develop accounts of the methodology of logic that are not straightforwardly realist, and check what are the implications of giving up the kind of realist approach that current anti-exceptionalist views have offered. Abandoning the unquestioned view that the aim of an epistemology for logic is to deliver the means to choose the true theory/theories is already a nice pointer towards a more productive engagement with science, according to antirealist lights. Here, we list some of the additional benefits that such an exploration could enjoy. 
\begin{itemize} 
\item We can overcome the uncritical view that there is a commonly accepted and successful description of scientific methodology, and that philosophers of logic are to blame for not having attempted to copy such a methodology before. The fact is that science itself is not a reign of consensus where the epistemology and metaphysics are worn on its sleeves; rather, consensus about the most important points appears only at the surface. Science is a very intricate human enterprise, and although its success is undeniable, it is also undeniable that the belief that it can deliver many instructive lessons as to how to achieve a similar level of results is not easily implemented (as the discussion on the metaphysics of science in the previous section  witnesses). 

\item Once one gives up the idea that the aim of logical inquiry is the search for the one true theory, the focus can shift to our \textit{understanding} of logic as it is, not as how philosophers would like it to be pursued. The fact is that logic is already a science, highly investigated and developed not only in philosophy departments, but also in mathematics and computer science departments, with its own research methods and research programs well established. Attempting to make the search for the one true logic as the major task for the epistemology of logic is at odds with current logical practice, at least when logicians are doing logic as a kind of investigation into the properties of logical systems. As \citet{Franks2015} has put it: 
\begin{quote}
One should warm up to the trend of identifying logic with the specialized scientific study of the relationships among various systems and their properties. This is, after all, how logicians use the word. Our preference to ignore questions about a logic's correctness stems not only from an interest in exploring the properties of possible logical systems in full generality but also from an appreciation, fostered by the study of logic, that no one such system can have all the properties that might be useful and interesting. \citep[163]{Franks2015}.
\end{quote}
By shifting to an antirealist way of pursuing the investigation of logic, instead of centering the whole epistemology in the search for the one true theory, we seem to end up much closer to current logical practice.\footnote{~That closeness to the practice is a virtue of epistemology for logic is also recognized by \citet{martin-2021}.} In the practice of logicians, questions of the correction of a system are of minor importance. Closely representing the practice of a science is seen in philosophy of science as a virtue of a philosophical account, while distance from practice is a negative point (as the debate between logical positivists and more practice-oriented approaches witness); the epistemology of logic could also learn this particular lesson.

\item Given the important role that classical logic plays in empirical science and classical mathematics, one could be led to think that this is the correct logic to be chosen as the true one \citep[see the discussion in][]{hjortland-2017,Hjortland2019,Bueno2017}. However, to recover a topic mentioned before, discussions related to the recapture of classical logic by non-classical theories in scenarios where classical logic is actually doing an important job should illustrate that there seems to be no fact of the matter that could help us to choose a logic for empirical science in general. As illustrated by Hjortland, in scenarios where one is willing to investigate the use of paracomplete logic in order to ground truth predicates and avoid semantic paradoxes, one needs not abandon classical mathematics:
\begin{quote}
But the theories nonetheless preserve classical mathematics. In model-theoretic terms, the fixed-point models of paracomplete theories are classical for the arithmetic language: no arithmetic sentence receives a nonclassical truth value. Importantly, the logical concepts of the paracomplete theories behave classically for sentences with classical truth values. As a result, the classical principles such the law of excluded middle are recaptured for arithmetic sentences and arguments. Of course, the presence of the truth predicate will complicate matters, but for good reason. The theory would be inconsistent if non-arithmetic sentences behaved classically. For the record, it is not only paracomplete theories that have this recapture property, but a number of nonclassical theories of truth (\textit{e.g.} paraconsistent theories, substructural theories). \citep[272]{Hjortland2019}.
\end{quote}
The result is that if science is to provide part of our evidence, as an anti-exceptionalist seems to believe, we seem to have to rely on considerations that are clearly not truth conducive (such as simplicity, familiarity, elegance, and so on) to guide our choices, in case we believe that a logic must be chosen. That is, of course, not a problem for an antirealist, but it clearly is an obstacle for those willing to apply abduction as a source of discovery of the one true logic. 

\item By rejecting the idea that there must be some kind of true/correct logic, the one that is responsive to the logical facts, we unburden our metaphysics. In fact, this does away with the very idea that there must be an (inaccessible) realm where logical consequence, validity itself, or \textit{logica ens} resides. Such an idea seems to be quite contrary to the ones guiding the investigation of logic and its diverse systems. Without requiring inaccessible logical facts, one brings logic closer to metaphysically deflationary accounts of science, and seems to account for some important views concerning the very nature of logical consequence. As Peter Smith has put it:
\begin{quote}
    If you think that there is [such a notion], start asking yourself questions like this. Is the intuitive notion of consequence constrained by considerations of relevance?---do ex falso quodlibet inferences commit a fallacy of relevance? When can you suppress necessarily true premises and still have an inference which is intuitively valid? What about the inference `The cup contains some water; so it contains some $H_2O$ molecules'? That necessarily preserves truth (on Kripkean assumptions): but is it valid in the intuitive sense?---if not, just why not? \citep[29]{Smith2011}.
\end{quote}
All of those enigmas disappear with the antirealist approach, counting as another benefit of the view. 

\item Consider the result of the realist approach to science and AEL regarding the issue of how many systems may be chosen as the correct one. Even though \citet{hjortland-2017} concedes as unattainable the idea that singling out a specific logic could be the final outcome of an abduction process, he still holds on to the idea that distinct logics may capture distinct regions of validity and this generates a pluralism of some sort. \citet{priest-2014} puts the issue in more neutral terms regarding the debate about monism versus pluralism: 
\begin{quote}
    I observe that this procedure does not prejudice the question of logical monism vs logical pluralism. If there is ``one true logic'' one's best appraisal of what this is is determined in the way I have indicated. If there are different logics for different topics, each of these is determined in the same way. Whether one single logic is better than many, is a ``meta-issue'', and is itself to be determined by similar considerations of rational theory-choice. \citep[216]{priest-2014}.
\end{quote}
As a result of giving up the belief that some system may be true, the debate between pluralism and monism in the philosophy of logic ends up deflated. This is certainly a topic to be explored, and the antirealist emphasis on understanding in opposition to truth clearly finds a place here as well. 
\end{itemize}

\section{Conclusion}\label{sec:5}

AEL is notoriously difficult to characterize, given the flexibility with which it has been characterized in the literature. However, taking into account the methodological agenda of AEL, which has been present in most of the discussion about AEL,  there is an underlying assumption concerning a new epistemology related to a methodology for theory choice, and also, the assumption that we will succeed in such a search for a new epistemology---ideally better than the traditional one---if we can choose the correct theory/theories among the many options available.

In this paper, we have sought to uncover these underlying assumptions and, given the anti-exceptionalist willingness to approach science, we have suggested that there are alternative ways to understand science that may be also promising for the anti-exceptionalist. That is, given that the lessons of science itself are not completely clear, the bridge between logic and empirical science is also not completely obvious. We have suggested that antirealist approaches to science also offer appropriate ways to understand science and logic, so that some of the difficulties plaguing current AEL may be understood as a result of using realist approaches to science. Although we have not explicitly defended any antirealist approach in particular, offering such additional possibilities is also part of the tools the antirealist has to understand science. We hope this sheds some light on further avenues that may be explored concerning AEL.  

\subsection*{Acknowledgments}
The order of authorship is alphabetical. The authors would like to thank: Evelyn Erickson for discussions on some of the topics explored in this paper; Sofia Abelha Meirelles for reading and commenting on an earlier version of this paper; the Philosophy of Logic reading group participants, hosted by the GFILL/CNPq Research Group and the Graduate Program in Philosophy of the Federal University of Maranhão. We also thank Sanderson Molick for the kind invitation for us to contribute to this volume with many authors we admire. This paper was developed while Raoni Arroyo was a visiting scholar at the Department of Philosophy, Communication, and Performing Arts, Roma Tre University, Rome, Italy, supported by grant \#2022/15992-8, São Paulo Research Foundation (FAPESP). Jonas R. B. Arenhart is partially supported by CNPq (Brazilian National Research Council). 

\printbibliography
\end{document}